\documentclass{bioinfo}
\copyrightyear{2005}
\pubyear{2005}

\usepackage{fixltx2e}

\begin{document}
\firstpage{1}

\title[]{Identifying edge clusters in networks via edge graphlet degree vectors (edge-GDVs) and edge-GDV-similarities}
  \author[]{Ryan W. Solava,\,$^{}$ Ryan P. Michaels,\,$^{}$ and
  Tijana Milenkovi\'{c}\,$^{}$\footnote{To whom correspondence should be
  addressed (tmilenko@nd.edu)}}
\address{$^{}$Department of Computer Science and Engineering,
  University of Notre Dame, IN 46556, USA}

\history{}

\editor{}

\maketitle

\begin{abstract}

Inference of new biological knowledge, e.g., prediction of protein
function, from protein-protein interaction (PPI) networks has received
attention in the post-genomic era.  A popular strategy has been to
cluster the network into functionally coherent groups of proteins and
predict protein function from the clusters.  Traditionally, network
research has focused on clustering of nodes.  However, why favor nodes
over edges, when clustering of edges may be preferred? For example,
nodes belong to multiple functional groups, but clustering of nodes
typically cannot capture the group overlap, while clustering of edges
can. Clustering of adjacent edges that share many neighbors was
proposed recently, outperforming different node clustering methods.
However, since some biological processes can have characteristic
``signatures'' throughout the network, not just locally, it may be of
interest to consider edges that are not necessarily adjacent.  Hence,
we design a sensitive measure of the ``topological similarity'' of
edges that can deal with edges that are not necessarily adjacent.  We
cluster edges that are similar according to our measure in different
baker's yeast PPI networks, outperforming existing node and edge
clustering approaches.

\end{abstract}

\section{Introduction}

A \emph{network} (\emph{graph}) consists of \emph{nodes} and
\emph{edges}. 
Network research spans many domains. Biomedical domain is no
exception.  We focus on \emph{protein-protein interaction (PPI)
networks}, in which nodes are proteins and undirected edges correspond
to physical binding between the proteins. Of all biological networks,
we focus on PPI networks since it is the proteins, gene products, that
carry out most biological processes, and they do so by interacting
with other proteins.  High-throughput screens for interaction
detection, such as yeast two-hybrid (Y2H) assays or affinity
purification coupled to mass spectrometry (AP/MS), have yielded
partial PPI networks for many model organisms and human
\citep{GiotSci03,Stelzl05,Yu2008,Simonis2009},
as well as for bacterial and viral pathogens
\citep{Parrish07,LaCount05}.  Many biological network
datasets are now publicly available \citep{HPRD,BIOGRID}.

Analogous to genomic sequence research, biological network research is
expected to have invaluable impacts on our biological understanding.
However, unlike genomic sequence research, biological network research
is in its infancy, owing to computational hardness of many graph
theoretic problems \citep{Cook1971}, as well as to incompleteness of
the available network data. Importantly, the number of functionally
uncharacterized proteins is large even for simple and well studied
model organisms \citep{Sharan2007}.
Functional characterization of proteins via computational analysis
could save resources needed for biological experiments.
In particular, PPI network analysis could help in suggesting top
candidates for future experimental validation, since proteins
aggregate to perform a function, and since PPI networks model these
aggregations.

Thus, it is no surprise that prediction of protein function
\citep{Sharan2007,Milenkovic2008} and the role of proteins in disease
\citep{Sharan2008,Radivojac2008,Goh2007,MMGP_Roy_Soc_09,Sharan10} from
PPI networks have received attention in the post-genomic era. For
example, it has been argued that proteins which are close in the
network are likely to be involved in similar biological processes
\citep{Sharan2007}, that ``topologically central'' proteins correspond
to ``biologically central'' (e.g., lethal, aging-, or cancer-related)
proteins \citep{Jeong01,Sharan2008,JonssonBates2006,Milenkovic2011},
or that proteins with similar topological neighbourhoods have similar
biological characteristics \citep{Milenkovic2008,Ho2010}.

A particularly popular strategy for functional characterization of
proteins has been to \emph{cluster} the network into functionally
``coherent'' groups of nodes and assign the entire cluster with a
function based on functions of its annotated members
\citep{Sharan2007,Sharan2008}.
A variety of clustering approaches exist, each with its own
(dis)advantages \citep{Brohee2006,Fortunato2010}.  Typically, they aim
to group nodes that are in a dense connected network region
\citep{Fortunato2010}.  Also, approaches exist that cluster
``topologically similar'' nodes without the nodes necessarily being
connected in the network.  This is important, since a biological
process can have characteristic topological ``signatures'' throughout
the network, not just localy in close network proximity
\citep{Milenkovic2008,MMGP_Roy_Soc_09,Ho2010}.  For example, we
designed a measure that computes the topological similarity of the
extended network neighborhoods of two nodes, without the nodes
necessarily being close in the network
\citep{Milenkovic2008}.  We found that 96\% of known cancer
gene pairs that are topologically similar according to our measure are
actually \emph{not} neighbours in the PPI network; instead, they are
at the shortest path distance of up to six \citep{MMGP_Roy_Soc_09}.
As such, they may be missed by approaches that focus on connected
nodes only.  We clustered proteins in the human PPI network that are
topologically similar and showed that function of a protein and its
network position are closely related \citep{Milenkovic2008} and that
the topology around cancer and non-cancer genes is different
\citep{MMGP_Roy_Soc_09}. We used these observations to predict new
cancer genes in melanogenesis-related pathways and our predictions
were validated phenotypically \citep{Ho2010}.

\begin{figure}[h!]
\begin{center}
  \resizebox{0.37\textwidth}{!}{\includegraphics[angle=0]{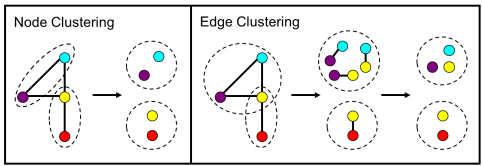}}
\caption{Node clustering (left) versus edge
  clustering (right).  } \label{node_edge_clustering}
\end{center} 
\end{figure}

Traditionally, network research has focused on clustering of
\emph{nodes} \citep{Fortunato2010}. However, a network consists of
nodes \emph{and} edges.  Hence, why favor nodes over edges, especially
when clustering of \emph{edges} may be preferred?  For example, since
nodes typically belong to multiple functional groups, and since
clusters are expected to correspond to the functional groups, it may
be desirable to allow for a node to belong to multiple clusters.
Clustering of nodes typically cannot capture the group overlap,
especially if the network is partitioned into disjoint node sets, as
is the case with many (although not all) node clustering approaches
\citep{Fortunato2010,Ahn2010}.  However, clustering of edges can
trivially capture the group overlap (Fig.
\ref{node_edge_clustering}).  Edge clustering methods were proposed
only recently \citep{Ahn2010,Evans2009}. Adjacent (connected) edges
that share many neighbors were defined as similar and were thus
clustered together (see below), outperforming different node
clustering methods, including a method which allows for the group
overlap \citep{Ahn2010}.  However, it may be of interest to consider
edges that are \emph{not} necessarily adjacent (see above).

Hence, we introduce a new measure of edge similarity that is not only
capable of dealing with edges that are not necessarily adjacent, but is
also a more sensitive measure of topology than the above
shared-neighborhood measure \citep{Ahn2010}.  For a fair evaluation of
our measure, when grouping edges that are similar according to our
measure, we precisely mimic the above edge clustering approach by
\cite{Ahn2010}. We show that using our measure results in clusters of
comparable or better quality.

\section{Approach}

We recently designed a graphlet-based measure of the topological
position of a \emph{node} in the network; graphlets are small
\emph{induced} subgraphs of a network (Fig. \ref{fig:graphlets})
\citep{Przulj07}.  This measure generalizes the degree of a
node that counts the number of edges that the node touches (where an
edge is the only 2-node graphlet) into the \emph{node graphlet degree
vector (node-GDV)} that counts the number of different graphlets that
the node touches, for all 2-5-node graphlets (Fig.
\ref{fig:graphlets}; also, see Methods). Hence, node-GDV describes the topology
of the node's up to 4-deep  neighborhood \citep{Milenkovic2008}.
This is effective: going to distance of four around a node captures a
large portion of the network due to the small-world nature of many
real networks \citep{Watts-Strogatz98}.  For this reason, and since
the number of graphlets on $n$ nodes increases exponentially with $n$,
using larger graphlets could unnecessarily increase the computational
complexity of the method.  Also, we designed
\emph{node-GDV-similarity} measure to compare node-GDVs of two nodes
and hence quantify the topological similarity of their extended
network neighborhoods \citep{Milenkovic2008}.

Since a graphlet consists of nodes \emph{and} edges, we now design
\emph{edge-GDV} to count the number of different 3-5-node graphlets
that an \emph{edge} touches (Fig. \ref{fig:graphlets}).  (We exclude
the count for the only 2-node graphlet, an edge, as each edge touches
exactly one 2-node graphlet, itself.)  Also, we design
\emph{edge-GDV-similarity} to compare edge-GDVs of two edges and hence
quantify the topological similarity of their extended network
neighborhoods.  Unlike the shared-neighborhood measure
\citep{Ahn2010}, edge-GDV-similarity can measure similarity between
edges independent on whether they are adjacent.  Also, by counting the
shared neighbors of end nodes of two (adjacent) edges, the
shared-neighborhood measure actually counts the 3-node paths that the
end nodes share \citep{Ahn2010}.  Since edge-GDV counts the different
3-5-node graphlets that an edge touches, including a 3-node path,
edge-GDV is a more constraining measure of topology. See Methods for
details.

We evaluate our approach against existing clustering methods, as
follows (also, see Methods). The existing edge clustering method
mentioned above, henceforth denoted by \emph{edge - shared
  neighborhood (edge-SN)}, was already shown to be superior to
different node clustering methods on four baker's yeast PPI networks
\citep{Ahn2010}.  For a fair evaluation, we mimic edge-SN exactly,
except that we use edge-GDV-similarity instead of the
shared-neighborhood measure as the distance metric for the same
clustering method, namely hierarchical clustering. Just as edge-SN, we
(initially) cluster only adjacent edges, and of all partitions, we
choose the one with the maximum density (see Methods).  Just as
edge-SN, we evaluate such partition with respect to four measures:
\emph{cluster coverage} (the portion of the network ``covered'' by
``non-trivial'' clusters), \emph{overlap coverage} (the amount of node
overlap between clusters), \emph{cluster quality} (enrichment of
clusters in Gene Ontology (GO) terms \citep{Go00}), and \emph{overlap
quality} (the correlation between the number of clusters and the
number of GO terms that nodes participate in). When applied to the
same yeast networks, our approach in comparable or superior to edge-SN
(and hence to the node clustering approaches that were outperformed by
edge-SN on the same networks). Thus, we gain by using a more sensitive
measure of topology compared to edge-SN.  Furthermore, when we cluster
both adjacent \emph{and} non-adjacent edges, our method in general
performs even better.  Hence, we gain additionally by using a measure
that can capture similarity of edges that are not necessarily
adjacent. We note that we do not propose a new edge clustering method
but a new edge similarity measure that can serve as a distance metric
for existing clustering methods.

\begin{methods}
\section{Methods}

\subsection{Data sets}\label{sect:data_sets}

We cluster the same four baker's yeast PPI networks that edge-SN was
evaluated on \citep{Ahn2010,Yu2008}: 1) \emph{Y2H} network, obtained
by Y2H, with 1,647 proteins and 2,518 PPIs; 2)
\emph{AP/MS} network, obtained by AP/MS, with 1,004 proteins and 8,319 PPIs;
3) \emph{LC} network, obtained by literature curation, with 1,213
proteins and 2,556 PPIs; and 4) \emph{ALL} network, representing the
union of Y2H, AP/MS, and LC, with 2,729 proteins and 12,174
PPIs. Using these different networks ensures that our method is robust
to different types of experiments for PPI detection.

\subsection{Related work}\label{sect:related_work}

We compare our method to three popular  node clustering
methods: clique percolation \citep{CliquePer2005}, greedy modularity
optimization \citep{NewmanGreedy2004}, and Infomap
\citep{Infomap2008}.  Also, we compare it to the existing edge
clustering algorithm, edge-SN \citep{Ahn2010}.  Briefly,
\emph{clique percolation} is the most prominent overlapping node
clustering algorithm, \emph{greedy modularity optimization} is the
most popular modularity-based technique, and \emph{Infomap} is often
considered the most accurate method available \citep{Ahn2010}.
\emph{Edge-SN} hierarchically groups adjacent edges whose non-common
end-nodes share many neighbors (see below).  We did not run 
these algorithms on the yeast networks ourselves. Instead, we use the
results reported by \cite{Ahn2010} who ran the algorithms on the same
networks. For details on how the methods were implemented, see
\cite{Ahn2010}.  We do explain how edge-SN
was implemented, as we implement our method in the same way (except
that we use a different distance metric).

Edge-SN algorithm works as follows.  If the set of node $i$ and its
neighbors is denoted as $n(i)$, the similarity between  adjacent
edges $e_{ik}$ and $e_{jk}$ with common node $k$ is $S(e_{ik}, e_{jk})
= | n(i) \cap n(j) | / | n(i) \cup n(j) |$. This
\emph{shared-neighborhood measure} is used as a distance metric for
single-linkage \emph{hierarchical clustering}.  With this method, a
tree, or dendrogram, is created.  Leaves of the tree are edges of the
network and an interior node in the tree represents a cluster made up
of all children of the node.  The tree is constructed by assigning
each edge to its own cluster and iteratively merging the most similar
pair of clusters.  The tree has to be cut in order to create a
partition of $K$ clusters.
To determine where to cut the tree, edge-SN uses an objective function
called \emph{partition density}, computed as follows.  For a network
with $M$ edges, $\{P_1, \cdots, P_K\}$ is a partition of the edges
into $K$ clusters.  Cluster $C$ has $m_C = |C|$ edges and $n_C = |
\cup_{e_{ij} \in C} {i,j}|$ nodes.  $C$'s density is $ D_C =
\frac{m_C - (n_C - 1)}{n_C(n_C - 1)/2 - (n_C - 1)} $,
and the partition density is $ D = \frac{2}{M} \sum_{K}m_C \frac{m_C -
(n_C - 1)}{(n_C - 2)(n_C - 1)} $.  For details, see \cite{Ahn2010}.
Edge-SN cuts the tree at different levels and chooses a partition with
the maximum value of $D$. However, meaningful structure may also exist
above and below the level corresponding to maximum $D$
\citep{Ahn2010}.

\subsection{New  measures of network topology: edge graphlet
  degree vector (edge-GDV) and edge-GDV-similarity}\label{sect:edge_GDV}

A graphlet is an \emph{induced} subgraph of graph $X$ that contains
\emph{all} edges of $X$ connecting its nodes (Fig.
\ref{fig:graphlets}).  We  generalized the degree of node $v$
that counts the number of edges that $v$ touches (where an edge is the
only 2-node graphlet, $G_0$ in Fig.  \ref{fig:graphlets}) into
\emph{node graphlet degree vector (node-GDV)} of $v$ that counts the
number of 2-5-node graphlets ($G_0$, $G_1$, $\ldots$, $G_{29}$ in Fig.
\ref{fig:graphlets}) that $v$ touches \citep{Milenkovic2008}. 
We need to distinguish between $v$ touching, e.g., a $G_1$ at an end
node or at the middle node, since $G_1$ admits an automorphism that
maps its end nodes to each other and the middle node to itself.  To
understand this, recall the following.  An isomorphism $f$ from graph
$X$ to graph $Y$ is a bijection of nodes of $X$ to nodes of $Y$ such
that $xy$ is an edge of $X$ if and only if $f(x)f(y)$ is an edge of
$Y$. An automorphism is an isomorphism from X to itself. The
automorphisms of $X$ form the automorphism group, $\mbox{Aut}(X)$. If
$x$ is a node of $X$, then the automorphism node orbit of $x$ is
$\mbox{Orb}(x) = \{ y \in V(X) | y = f(x) \mbox{ for some } f \in
\mbox{Aut}(X)\}$, where $V(X)$ is the set of nodes of $X$.  Thus, end
nodes of a $G_1$ belong to one node orbit, while its middle node
belongs to another one.  There are 73 node orbits for 2-5-node
graphlets.  Hence, node-GDV of $v$ has 73 elements counting how many
node orbits of each type touch $v$ ($v$'s degree is the first
element).  It captures $v$'s up to 4-deep neighborhood and thus a
large portion of real networks, as they are small-world
\citep{Watts-Strogatz98}.

Since a graphlet contains nodes \emph{and} edges, we propose a new
graphlet-based measure of the topological position of an \emph{edge}
in the network. We define \emph{edge-GDV} to count the number of
graphlets that an \emph{edge} touches at a given ``edge orbit'' (Fig.
\ref{fig:graphlets}).  We define edge orbits are follows.  Given the
automorphism group of graph $X$, $\mbox{Aut}(X)$, if $xy$ is an edge
of $X$, then the edge orbit of $xy$ is $\mbox{Orb}_e(xy) = \{ zw \in
E(X) | z = f(x) \mbox{ and } w = f(y) \mbox{ for some } f \in
\mbox{Aut}(X)\}$, where $E(X)$ is the set of edges of $X$.  For
example, in Fig.  \ref{fig:graphlets}, in a $G_1$, both edges are in
edge orbit 1. In a $G_2$, all three edges are in edge orbit 2. In a
$G_3$, the two ``outer'' edges are in edge orbit 3, while the
``middle'' edge is in edge orbit 4. And so on. There are 68 edge
orbits for 3-5-node graphlets. (We intentionally exclude orbit 0 in
the only 2-node graphlet, $G_0$, as each edge touches exactly one
$G_0$, namely itself.)

\begin{figure}[h!]
\begin{center}
\resizebox{0.485\textwidth}{!}{\includegraphics[angle=0]{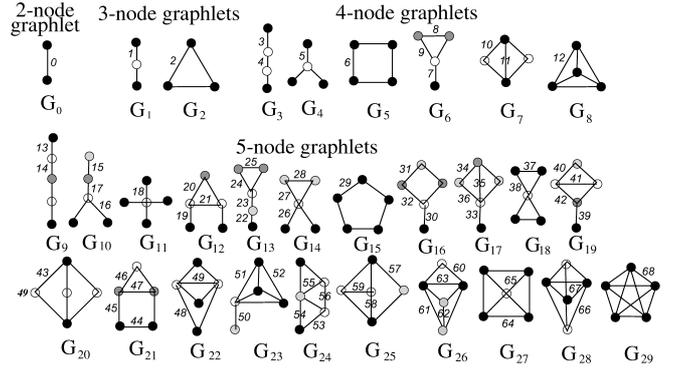}}\\
\end{center} 
\begin{center}
\caption{All the connected graphs on 2 to 5 nodes.
  When appearing as induced subgraphs of a network, they are called
  {\it graphlets}.  They contain 73 topologically unique node types,
  or ``node orbits.''  In a graphlet, nodes in the same node orbit are
  of the same shade \citep{Przulj07}. They also contain 69
  topologically unique edge types, or ``edge orbits.''  (3-5-node
  graphlets contain 68 edge orbits.) Edge orbit of an edge is defined
  by node orbits of its end nodes. In a graphlet, different edge
  orbits are numbered differently.  } \label{fig:graphlets}
\end{center} 
\end{figure}

Comparing edge-GDVs of two edges gives a sensitive measure of their
topological similarity, since their \emph{extended} network
neighborhoods are compared. Using some existing measure, e.g.,
Euclidean distance, to compare edge-GDVs might be inappropriate, as
some edge orbits are not independent.  Instead, we design {\it
edge-GDV-similarity} measure as follows.  For an edge $e$, $e_i$ is
the $i^{th}$ element of its edge-GDV.  The distance between the
$i^{th}$ edge orbits of edges $e$ and $f$ is $D_i(e,f) = w_i \times
\frac{|log(e_i + 1) - log(f_i + 1)|}{log(max\{e_i, f_i\} + 2)}$, where
$w_{i}$ is the weight of edge orbit $i$ that accounts for edge orbit
dependencies.  For example, the differences in counts of orbit 2 of
two edges will imply the differences in counts of all other orbits
that contain orbit 2, such as orbits 8-12 (Fig.
\ref{fig:graphlets}). This is applied to all edge orbits: the smaller
the number of orbits that affect orbit $i$ (including itself),
$o_{i}$, the higher its weight $w_i$, where $w_i = 1 -
\frac{log(o_i)}{log(68)}$. Clearly, $w_i$ is in (0,1] and the highest
weight of $1$ is assigned to orbit $i$ with $o_i=1$.  The $log$ is
used in the formula for $D_i$ because the $i^{th}$ elements of two
edge-GDVs can differ by several orders of magnitude and we do not want
the distance between edge-GDVs to be dominated by large values; also,
we want to account for the relative difference between $e_i$ and $f_i$
and that is why we divide by the value of the denominator, which also
scales $D_{i}$ to [0, 1).  The constants are added to prevent $D_i$ to
be infinite.  The total distance is $D(e,f) =
\frac{\sum_{i=0}^{68}D_i}{\sum_{i=0}^{68}w_i}$.  Finally,
edge-GDV-similarity is $S(e,f) = 1 - D(e,f)$.  It is in (0, 1].  The
higher the edge-GDV-similarity, the higher the topological similarity
of edges' extended network neighborhoods.  We design
edge-GDV-similarity as described because we already designed
node-GDV-similarity, which compares node-GDVs, in a similar way
\citep{Milenkovic2008}, and because we showed in different contexts
that node-GDV-similarity successfully extracts function from network
topology
\citep{HGRAAL,Memisevic10b,GRAAL,MMGP_Roy_Soc_09,Milenkovic2011}.  So,
we expect edge-GDV-similarity to successfully extract function from
topology as well.

\subsection{Our clustering strategies}\label{sect:edge_clustering}

We cluster the yeast PPI networks in the same manner as edge-SN,
except that we use edge-GDV-similarity as the distance metrics instead
of using the shared-neighborhood measure. Initially, for a fair
comparison with edge-SN, we cluster adjacent edges only, to test if
and how much we gain by using our more sensitive measure of edge
similarity. Later on, we cluster all edges, to test if and how much we
gain by taking into account edges that are not necessarily adjacent.
Some further information is provided below, after defining measures of
partition quality.

\subsection{Quality of partitions}\label{sect:quality_of_clusters}

We evaluate a partition with respect to the same measures that were
used by edge-SN: cluster coverage (CC), overlap coverage (OC), cluster
quality (CQ), and overlap quality (OQ).  CC is the fraction of nodes
that belong to at least one ``non-trivial'' cluster of three or more
nodes.  OC is the average number of non-trivial clusters that nodes
belong to.  CQ is the ratio of the average Gene Ontology (GO) term
\citep{Go00} similarity  over all node pairs that are in
at least one same cluster and the average GO terms similarity over all
node pairs in the network.  OQ is the mutual information between the
number of GO terms and the number of non-trivial clusters that
proteins are involved in.  Raw values for the four measures do not
necessarily fall in $[0, 1]$.  Hence, just as \cite{Ahn2010}, we
normalize each measure such that the best method has a value of
one. Then, the overall partition quality is the sum of these four
normalized measures, such that the maximum achievable score is four.

We can now note the following.  To mimic \cite{Ahn2010}, we would
report for a given network the partition with maximum partition
density $D$. However, we find that CC is strongly negatively
correlated with CQ and OQ, and sometimes with OC, over all of our
partitions (Fig. \ref{fig:correlations}). Thus, choosing the partition
with low CC would result in high CQ and OQ (and sometimes OC), hence
artificially increasing the overall partition quality. Since in three
out of four yeast networks CC is lower for edge-SN than for the node
clustering methods, it might not be surprising that edge-SN's overall
partition quality is the highest. Analogously, since edge-SN's
partitions with maximum $D$ have lower CC than our partitions with
maximum $D$, our partitions may have lower overall partition quality
simply because of the strong negative correlation between CC and other
measures.  Hence, we find the partition with maximum $D$ among all
partitions that have CC less than or equal to CC of edge-SN's
partition with maximum $D$. Then, we report either the partition
obtained in this way or the partition with maximum $D$ (independent of
its CC), whichever has better overall partition quality. Furthermore,
when we cluster both adjacent and non-adjacent edges, selecting the
partition based on its density, as just described, might be
inappropriate (see above).  Thus, when we cluster both adjacent and
non-adjacent edges, we also report the partition with the best overall
partition quality.

\begin{figure*}[tbp]
\begin{center}
  \textbf{(A)}\resizebox{0.22\textwidth}{!}{\includegraphics[angle=0]{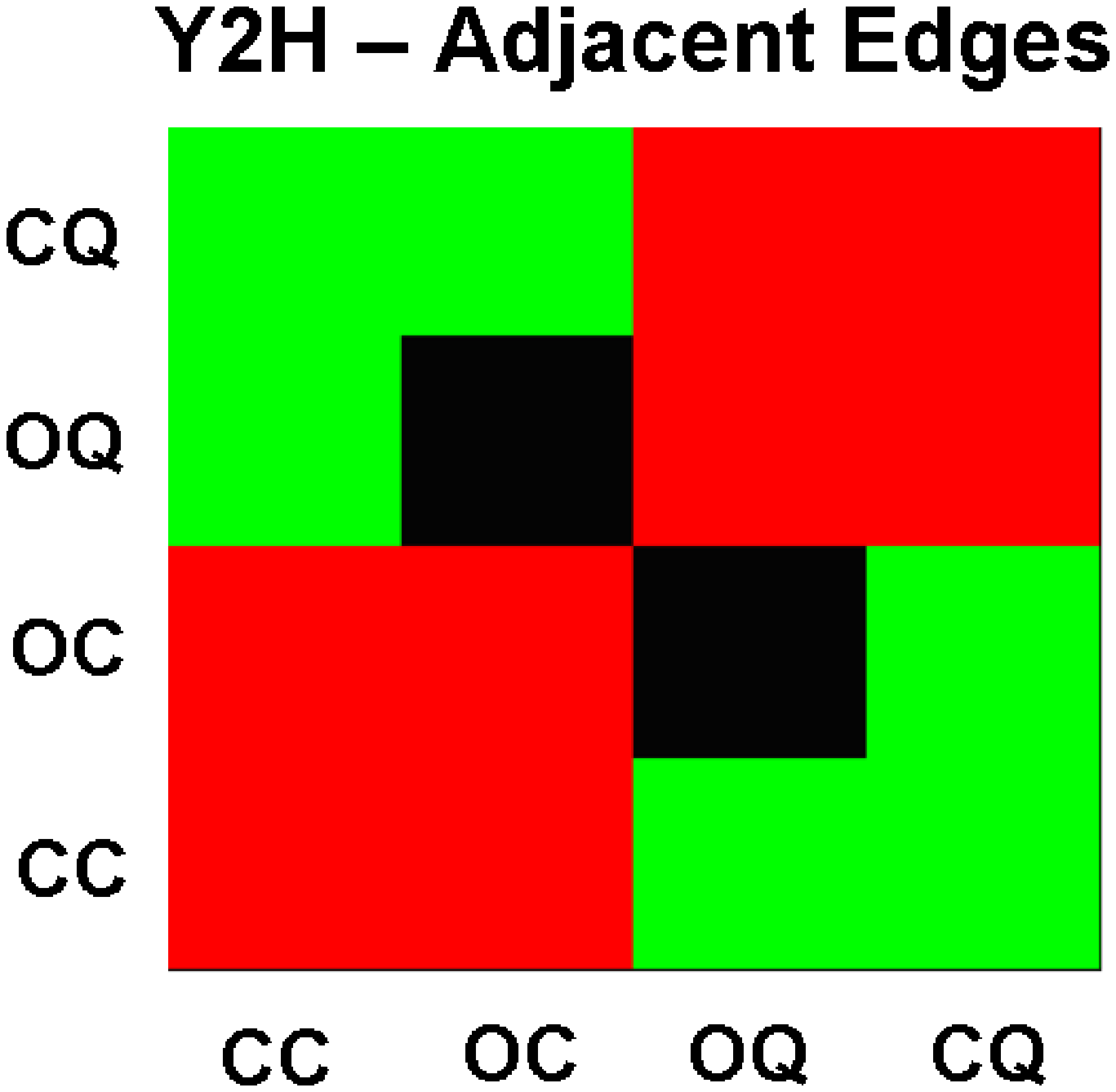}}
  \textbf{(B)}\resizebox{0.22\textwidth}{!}{\includegraphics[angle=0]{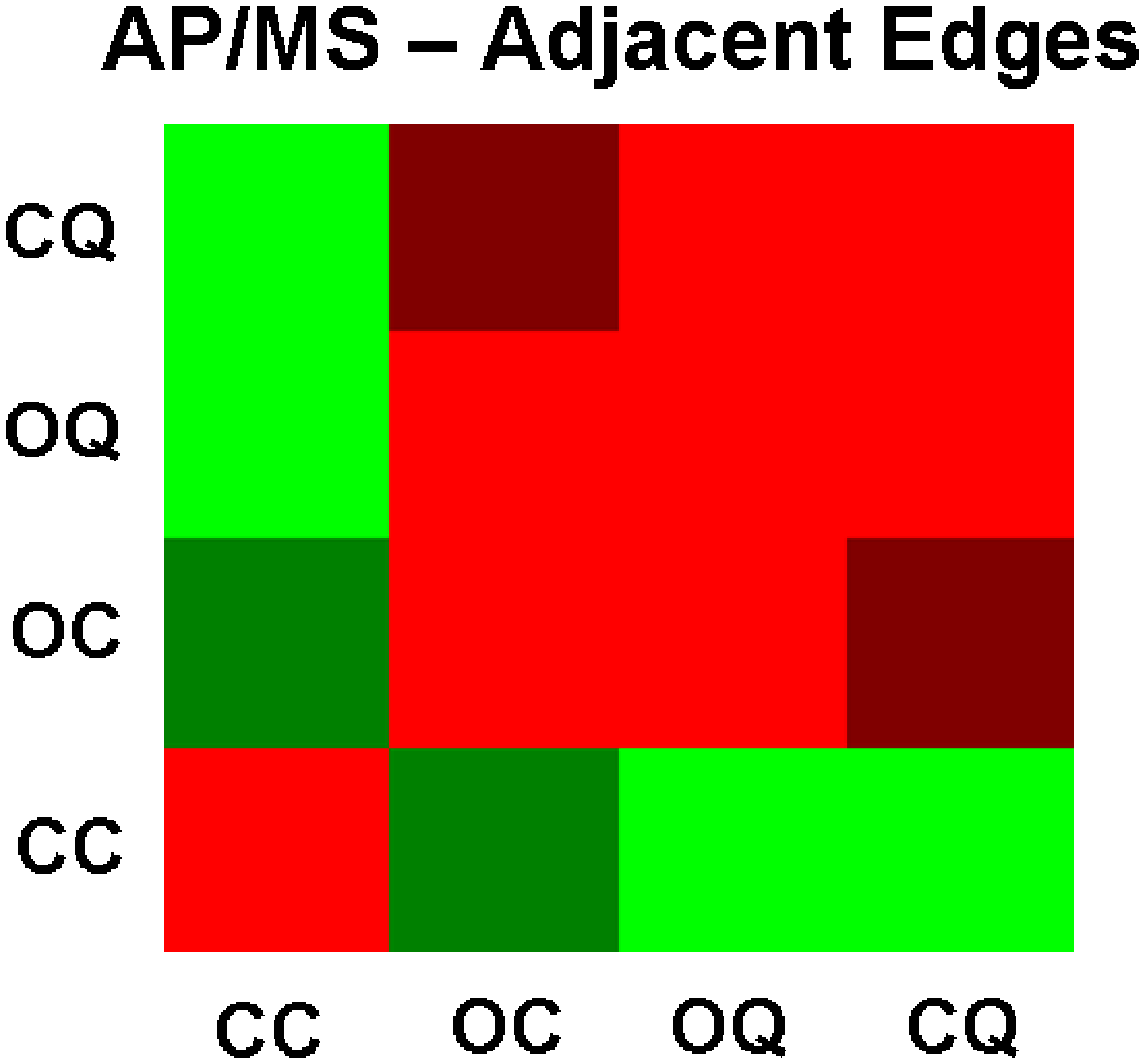}}
  \textbf{(C)}\resizebox{0.22\textwidth}{!}{\includegraphics[angle=0]{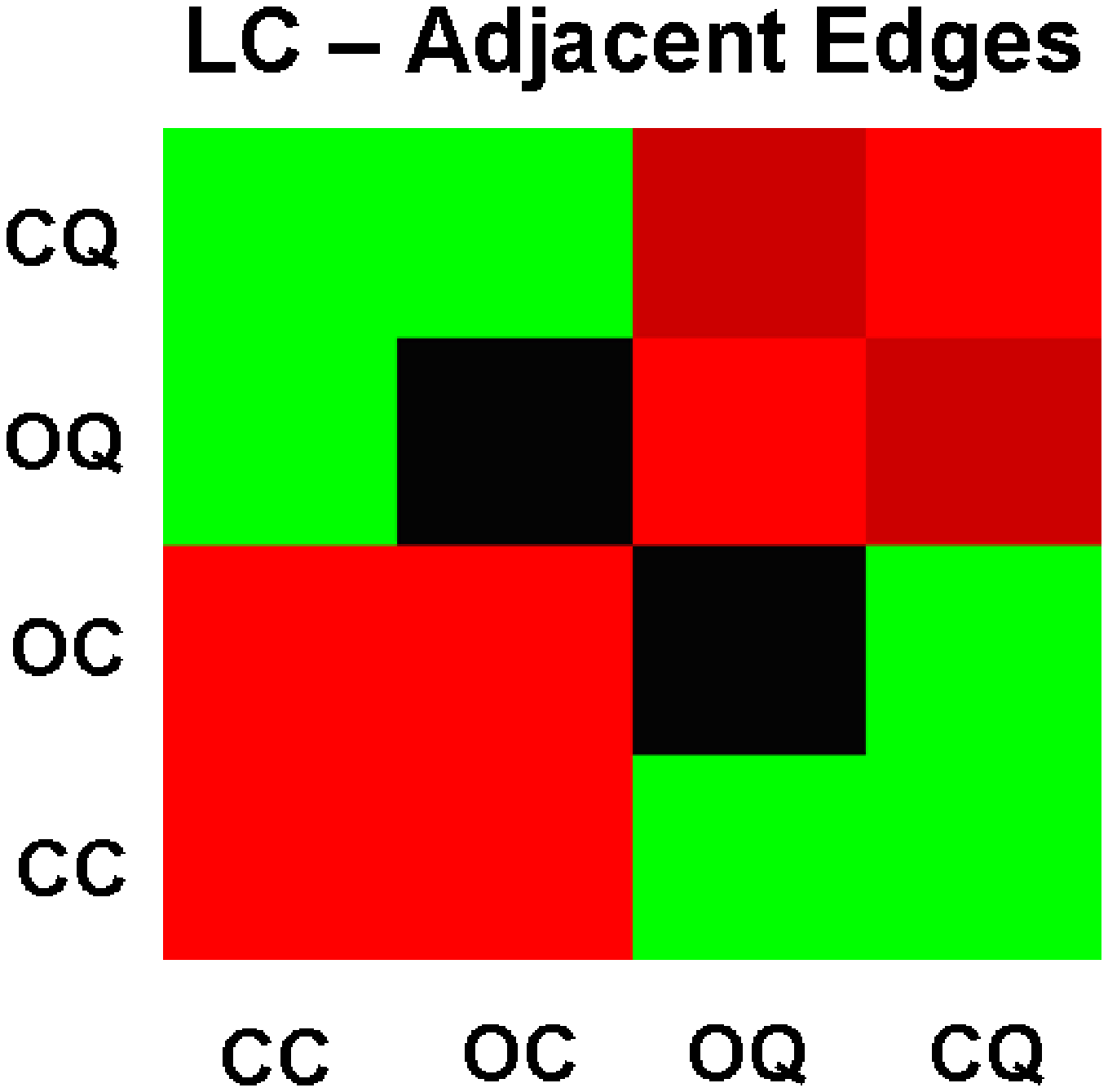}}
  \textbf{(D)}\resizebox{0.22\textwidth}{!}{\includegraphics[angle=0]{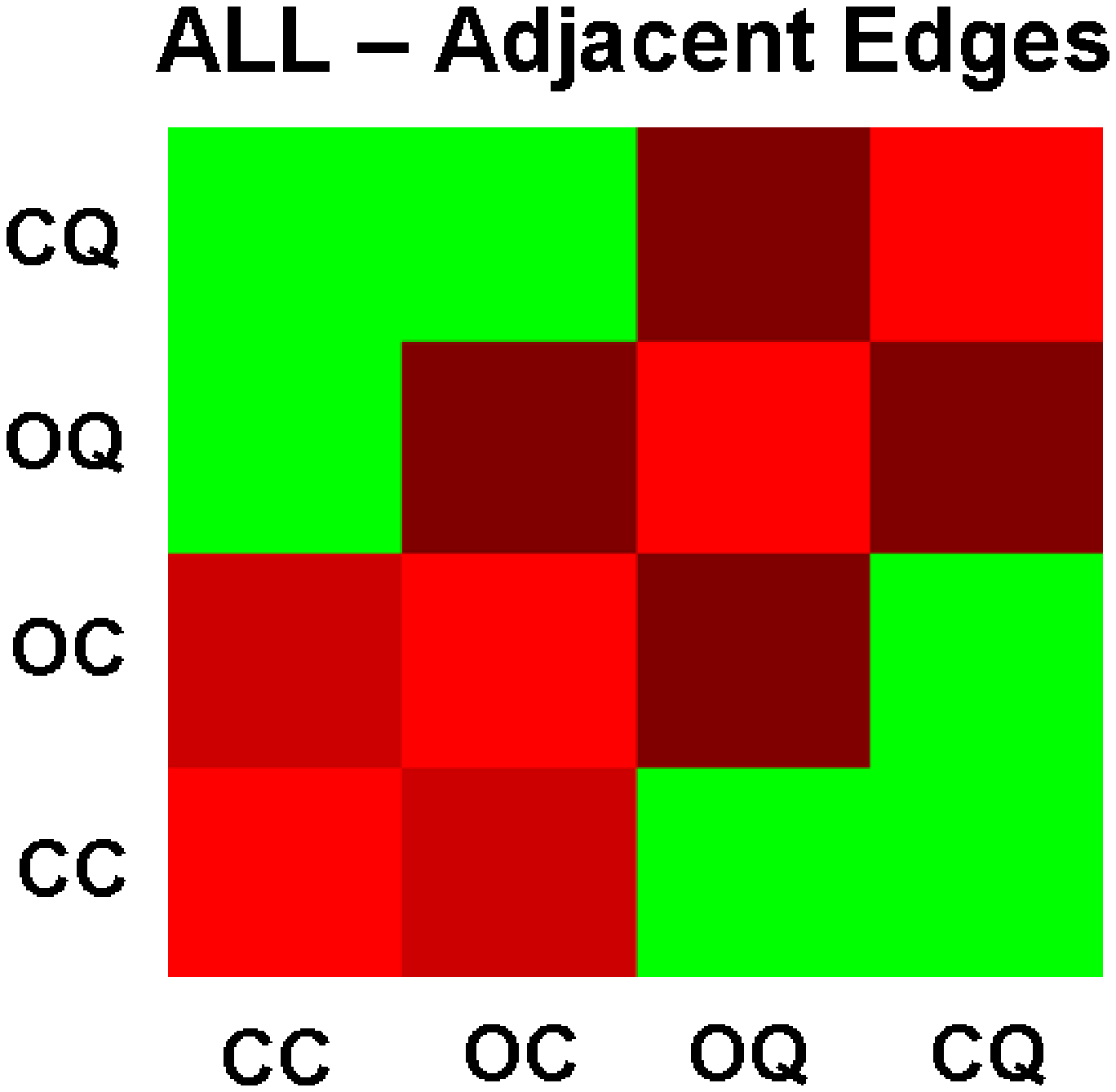}}\\
  \textbf{(E)}\resizebox{0.22\textwidth}{!}{\includegraphics[angle=0]{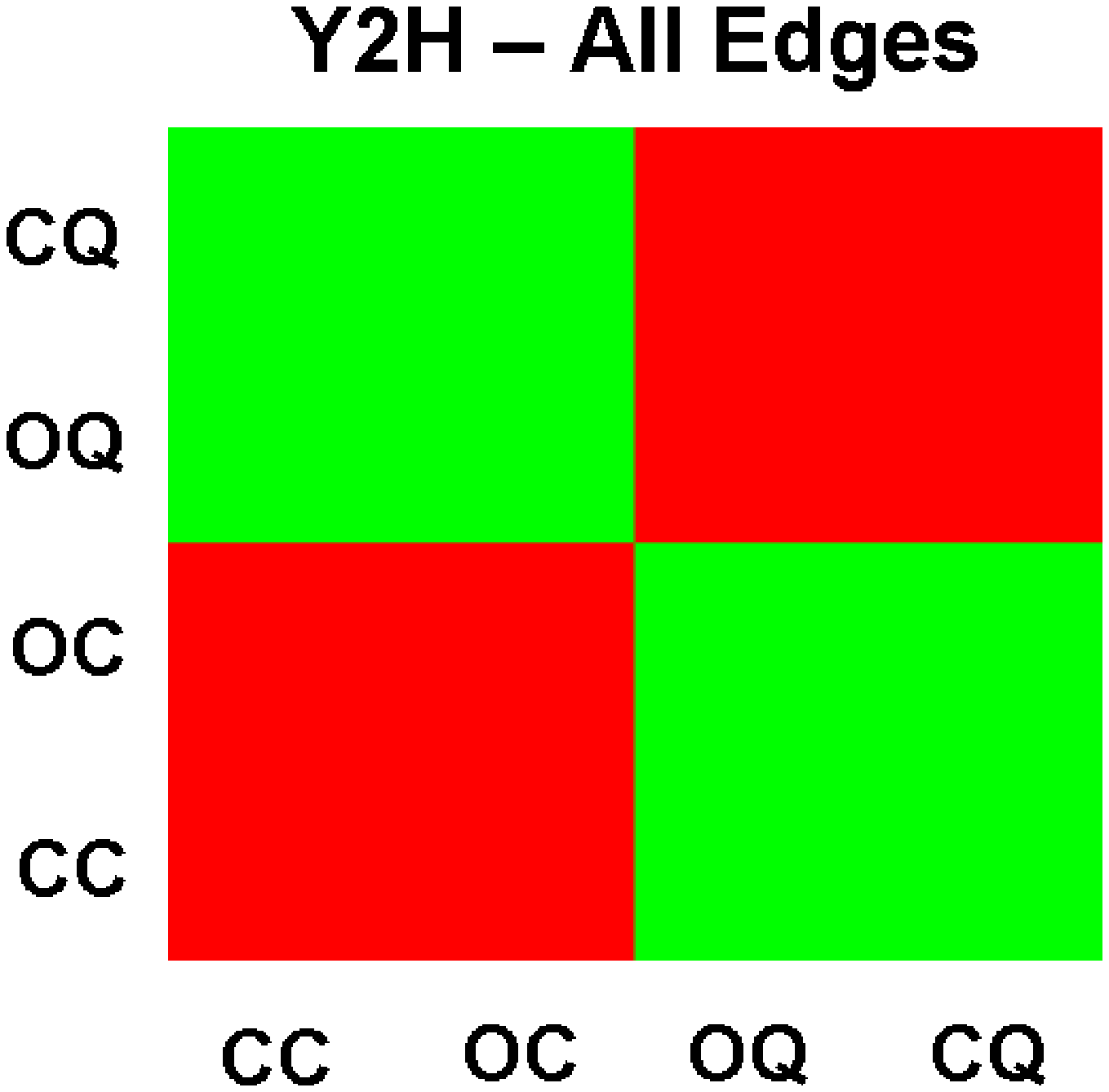}}
  \textbf{(F)}\resizebox{0.22\textwidth}{!}{\includegraphics[angle=0]{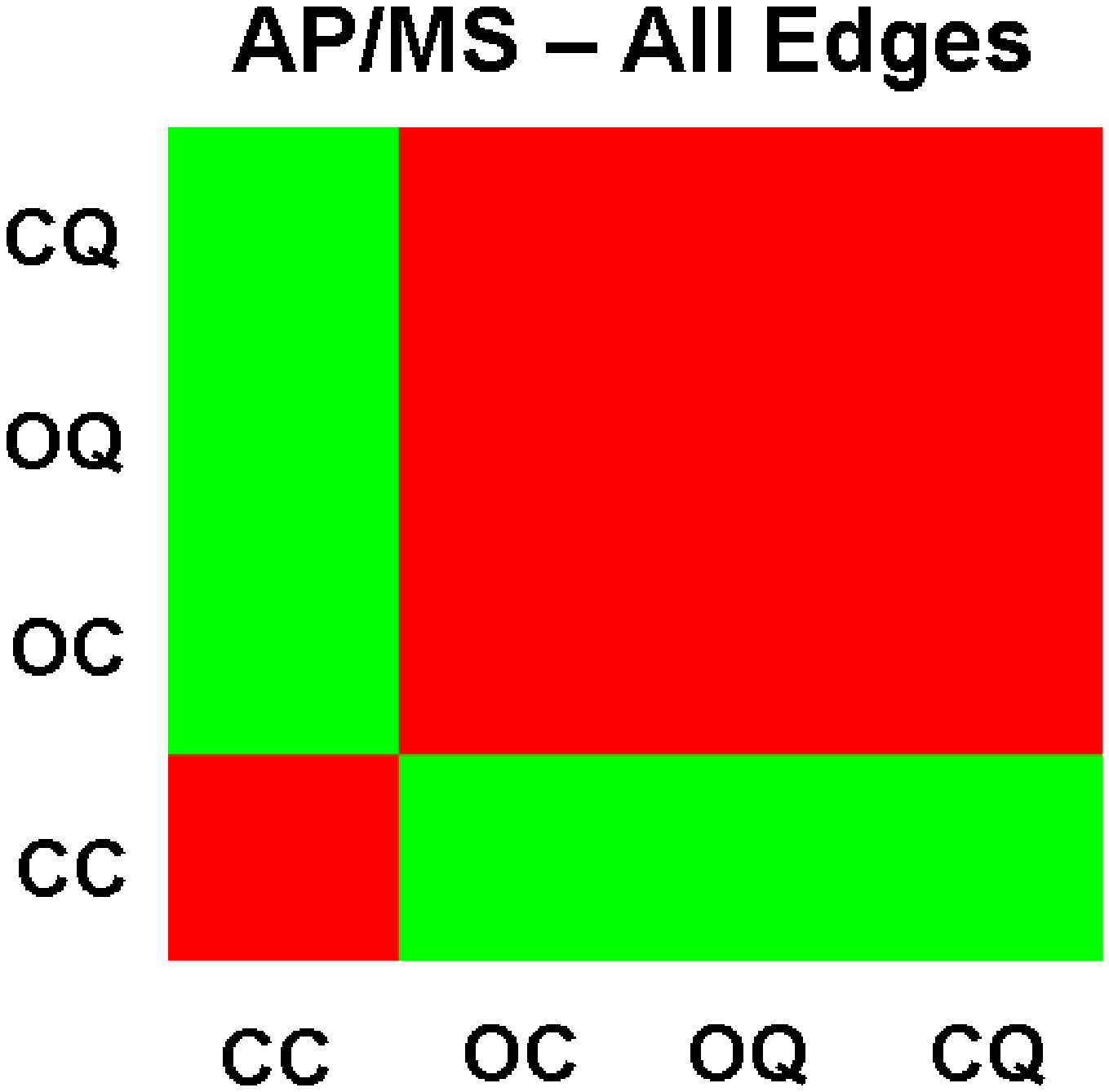}}
  \textbf{(G)}\resizebox{0.22\textwidth}{!}{\includegraphics[angle=0]{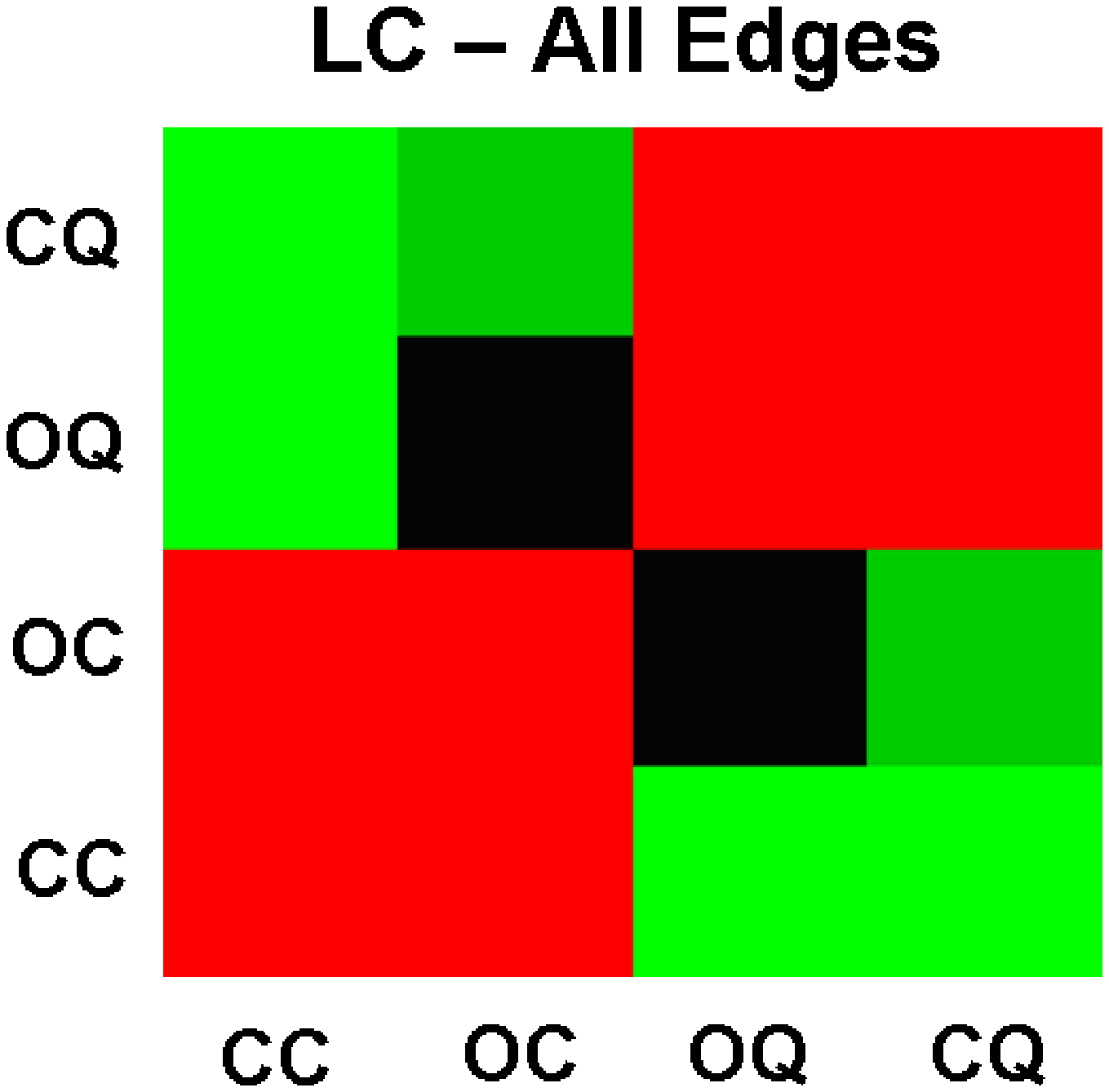}}
  \textbf{(H)}\resizebox{0.22\textwidth}{!}{\includegraphics[angle=0]{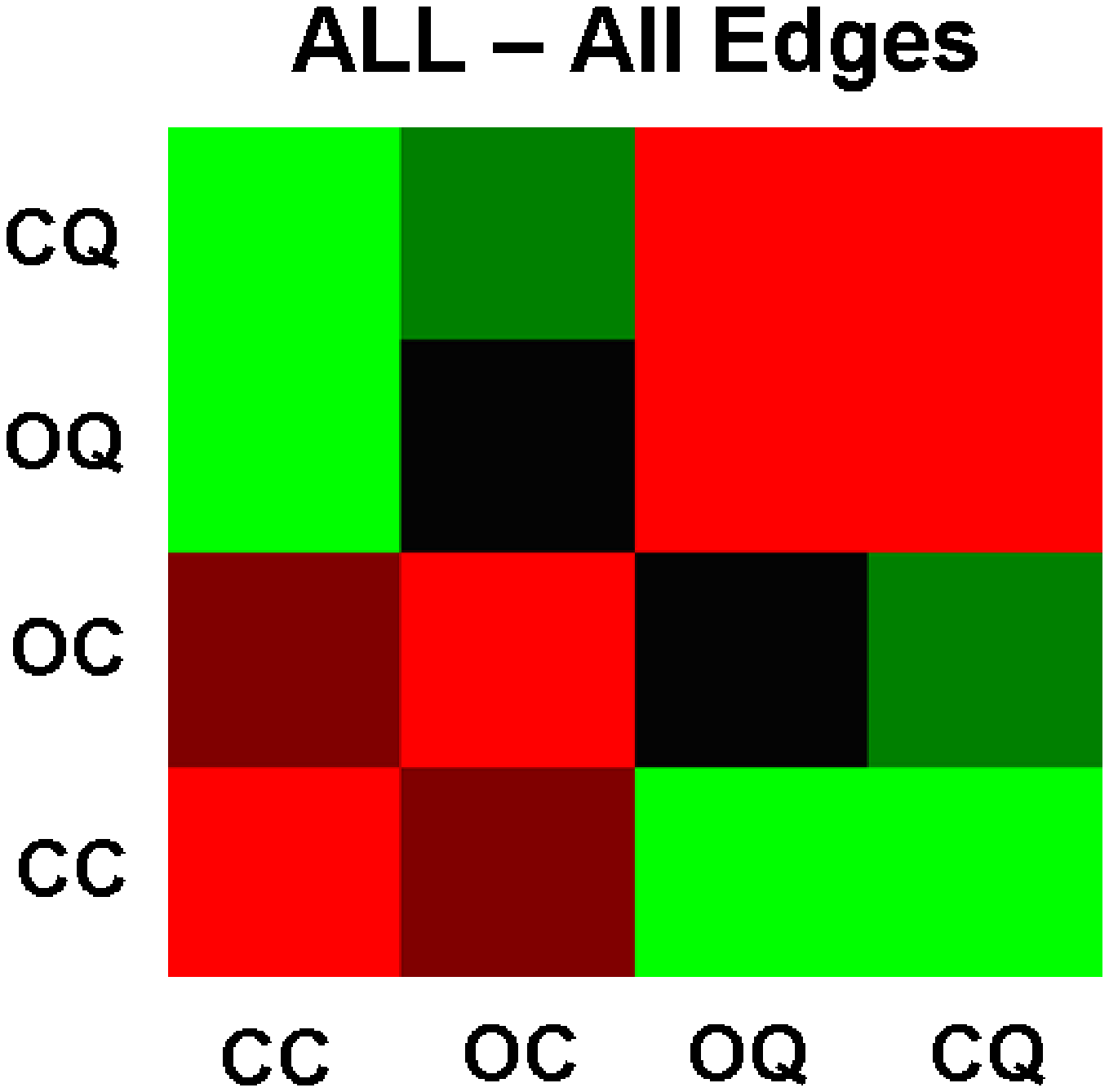}}\\
  \resizebox{0.32\textwidth}{!}{\includegraphics[angle=0]{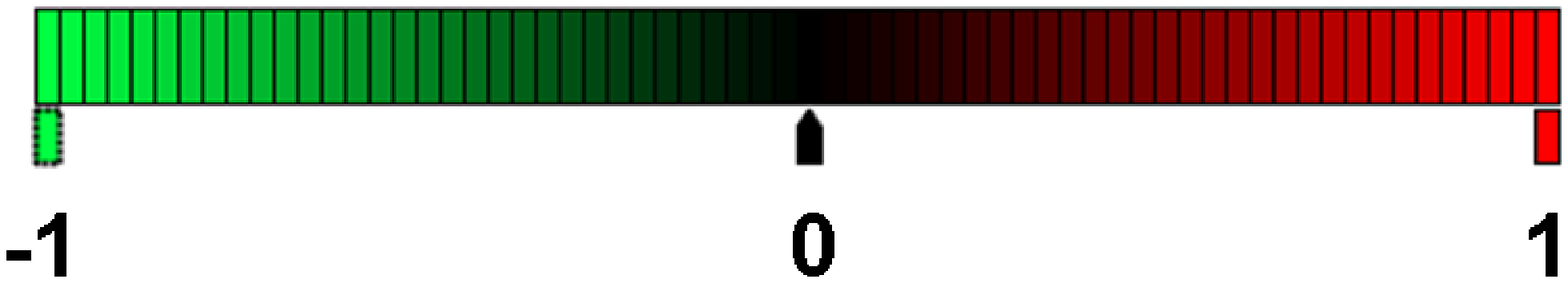}}
\caption{Pairwise Pearson correlations for four
  measures of partition quality (cluster coverage -- CC, overlap
  coverage - OC, overlap quality -- OQ, and cluster quality -- CQ)
  over all partitions for a given yeast PPI network (Y2H, AP/MS, LC,
  and ALL), when clustering adjacent edges only (Adjacent Edges;
  panels \textbf{A-D}) and when clustering both adjacent and
  non-adjacent edges (All Edges; panels \textbf{E-H}). In all eight
  panels, CC is strongly negatively correlated with OQ and CQ. }
  \label{fig:correlations}
\end{center} 
\end{figure*}

\end{methods}

\section{Results}

\begin{figure*}[h!]
\begin{center}
  \textbf{(A)}\resizebox{0.45\textwidth}{!}{\includegraphics[angle=0]{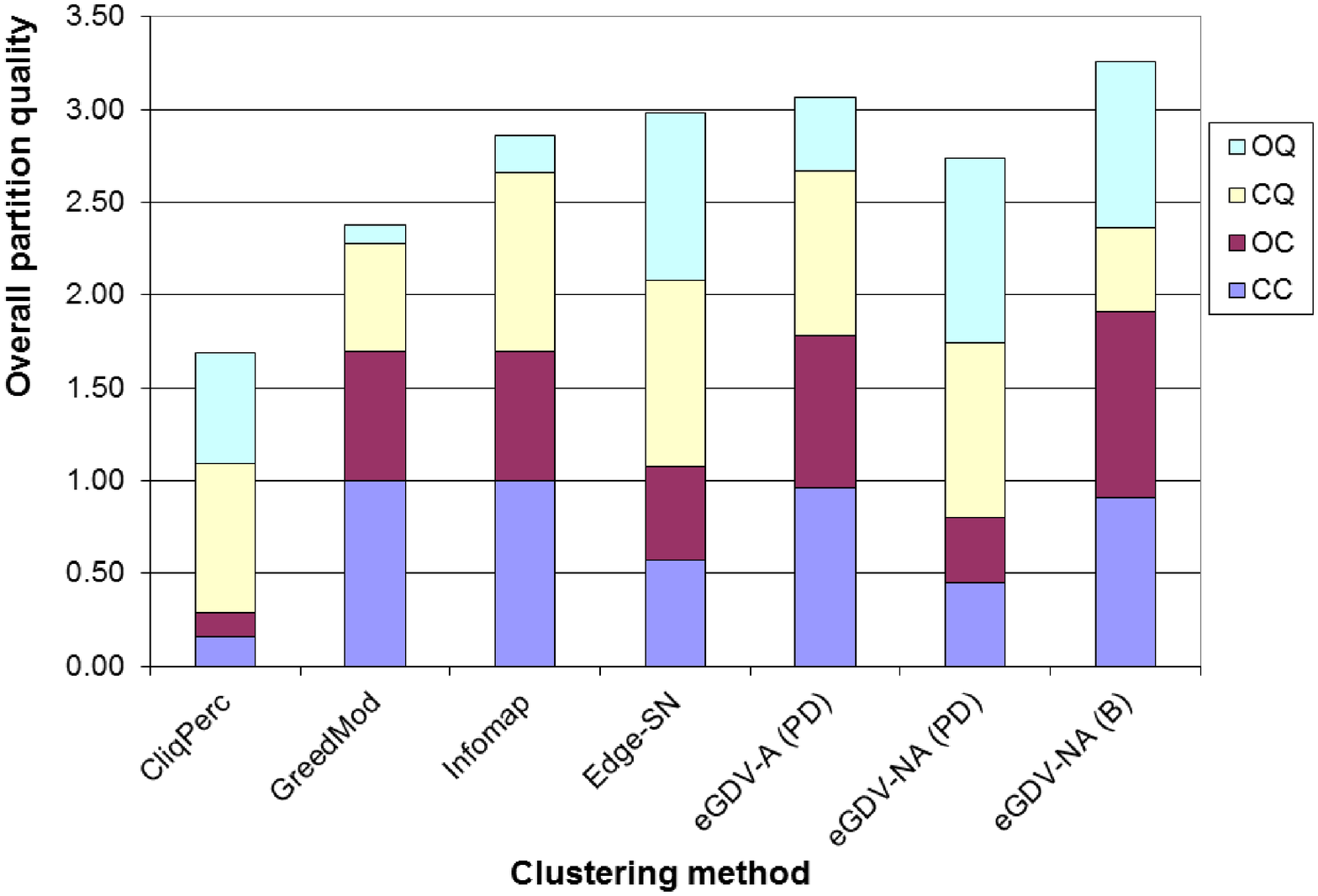}}
 \textbf{(B)} \resizebox{0.45\textwidth}{!}{\includegraphics[angle=0]{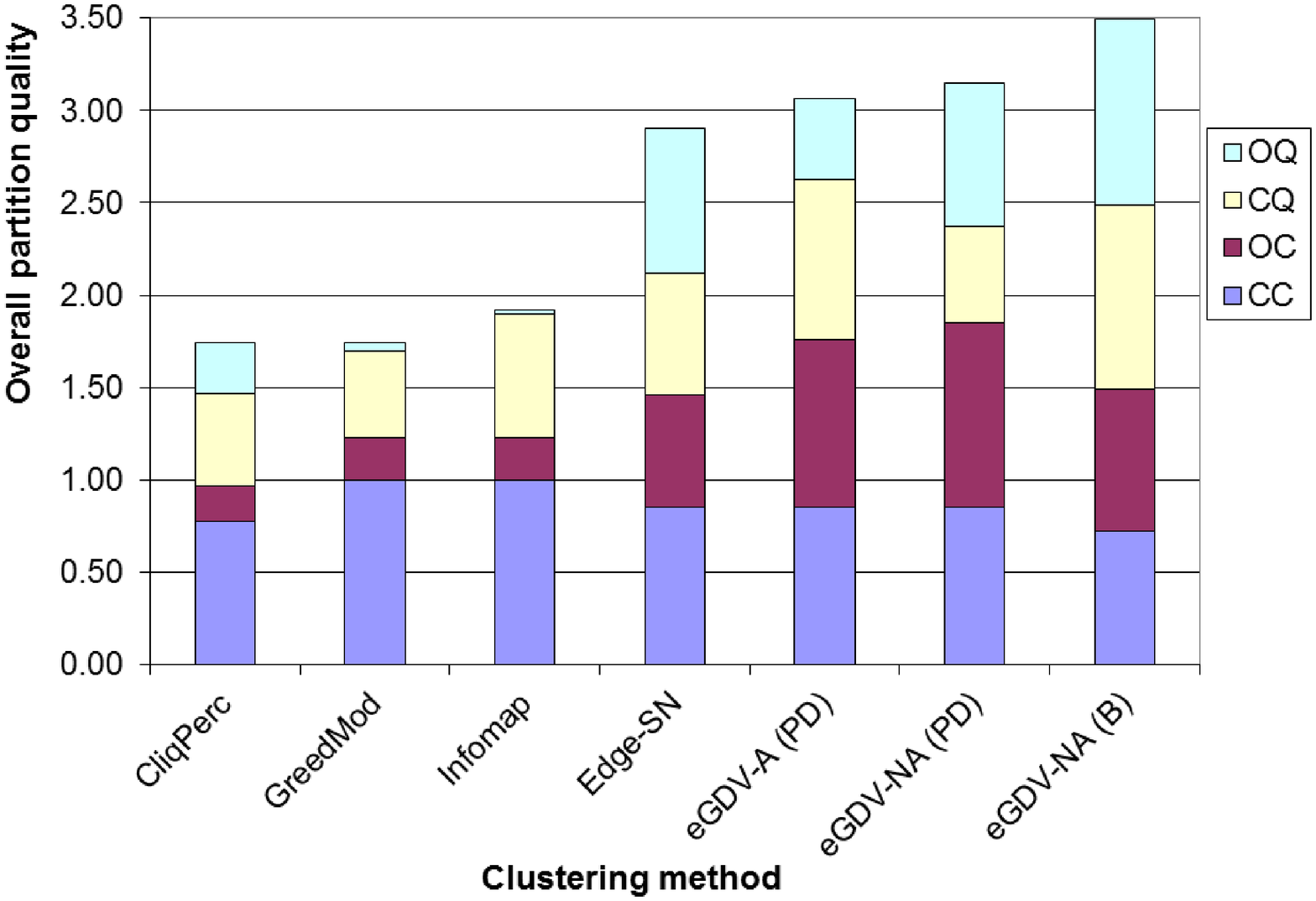}}\\
 \textbf{(C)} \resizebox{0.45\textwidth}{!}{\includegraphics[angle=0]{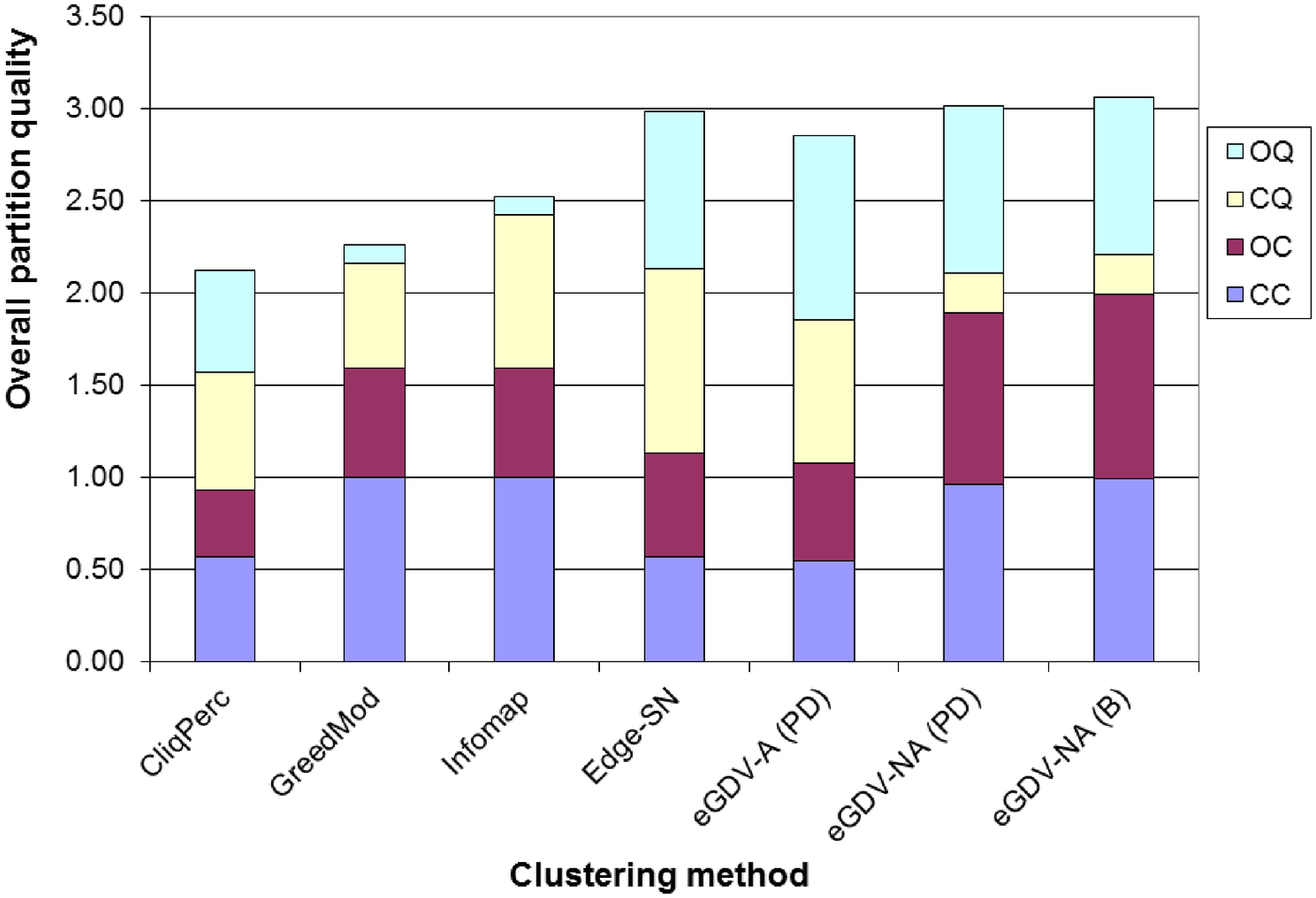}}
 \textbf{(D)} \resizebox{0.45\textwidth}{!}{\includegraphics[angle=0]{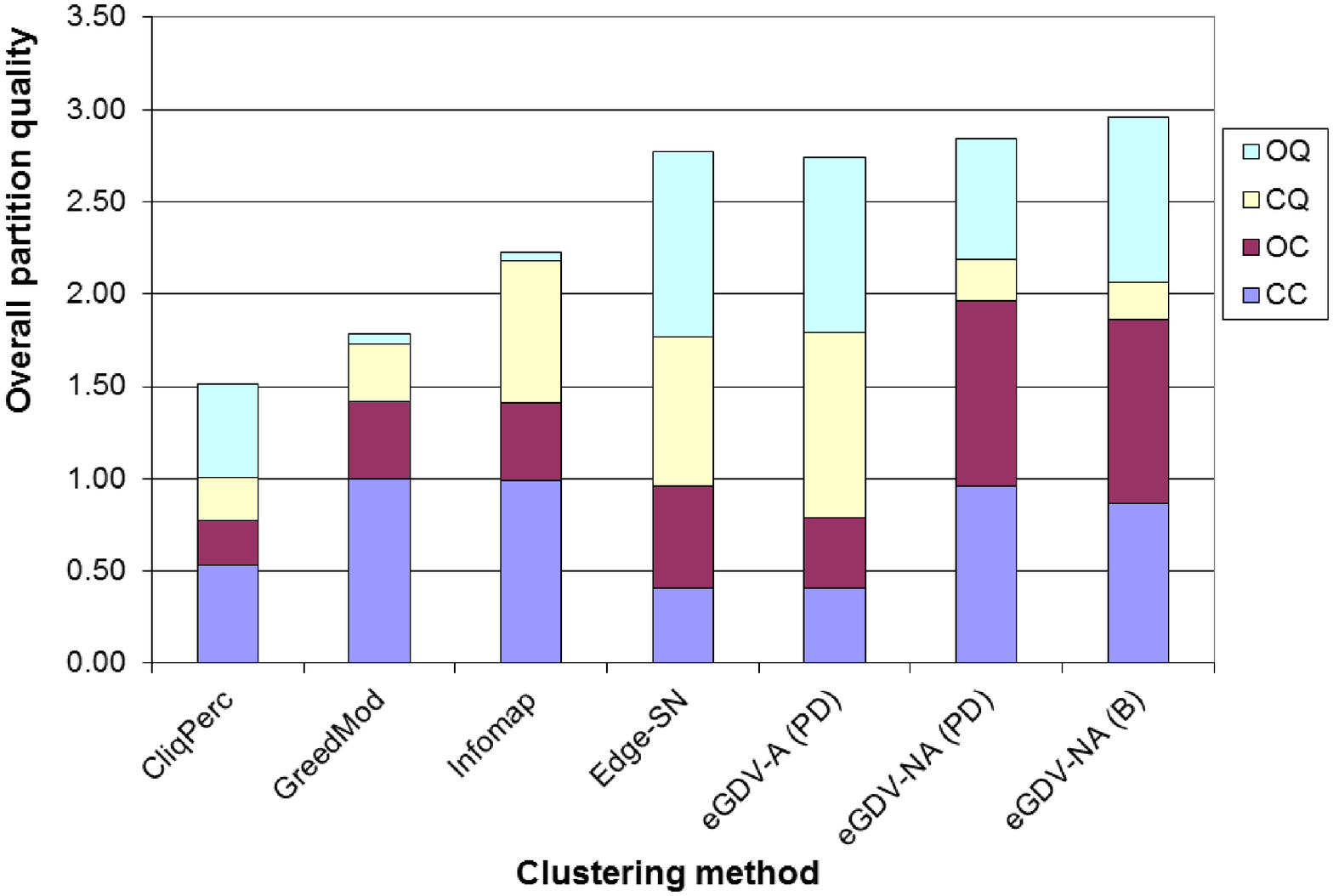}}
\caption{Method comparison for \textbf{(A)} Y2H, \textbf{(B)}  AP/MS, \textbf{(C)} LC, and \textbf{(D)} ALL  yeast PPI networks. 
  The following methods are compared: clique percolation (CliqPerc),
  greedy modularity optimization (GreedMod), Infomap, edge-SN, our
  method when clustering adjacent edges only and choosing the
  partition with the maximum density (eGDV-A-D), our method when
  hierarchically clustering both adjacent and non-adjacent edges and
  choosing the partition with the maximum density (eGDV-NA-PD), and
  our method when hierarchically clustering both adjacent and
  non-adjacent edges and choosing the partition with the best overall
  partition quality (eGDV-NA-B). Clustering methods are compared with
  respect to the following measures: cluster coverage (CC), overlap
  coverage(OC), cluster quality (CQ), and overlap quality (OQ). The
  overall partition quality score ($y$-axis) is the sum of these four
  measures after each is normalized to [0,1], such that the maximum
  achievable score is four.  }
\label{fig:comparison1}
\end{center} 
\end{figure*}

We evaluate our method against three node clustering methods (clique
percolation -- CliqPerc, greedy modularity optimization -- GreedMod,
and Infomap) and one edge clustering method (edge-SN) on four yeast
PPI networks (Y2H, AP/MS, LC, and ALL), with respect to four measures
of partition quality (cluster coverage -- CC, overlap coverage - OC,
cluster quality -- CQ, and overlap quality -- OQ) that are combined
into the normalized overall partition quality; see Methods.  We denote
our method when clustering \textbf{a}djacent edges only and reporting
the partition with the maximum \textbf{d}ensity as eGDV-A-D.  We
denote our method when clustering both adjacent and
\textbf{n}on-\textbf{a}djacent edges and reporting the partition with
the maximum \textbf{d}ensity as eGDV-NA-D.  We denote our method when
clustering both adjacent and \textbf{n}on-\textbf{a}djacent edges and
reporting the partition with the \textbf{b}est overall partition
quality as eGDV-NA-B.  See Methods for details. Results are shown in
Fig.
\ref{fig:comparison1}. 

\begin{figure*}[htbp!]
\begin{center}
  \resizebox{0.5\textwidth}{!}{\includegraphics[angle=0]{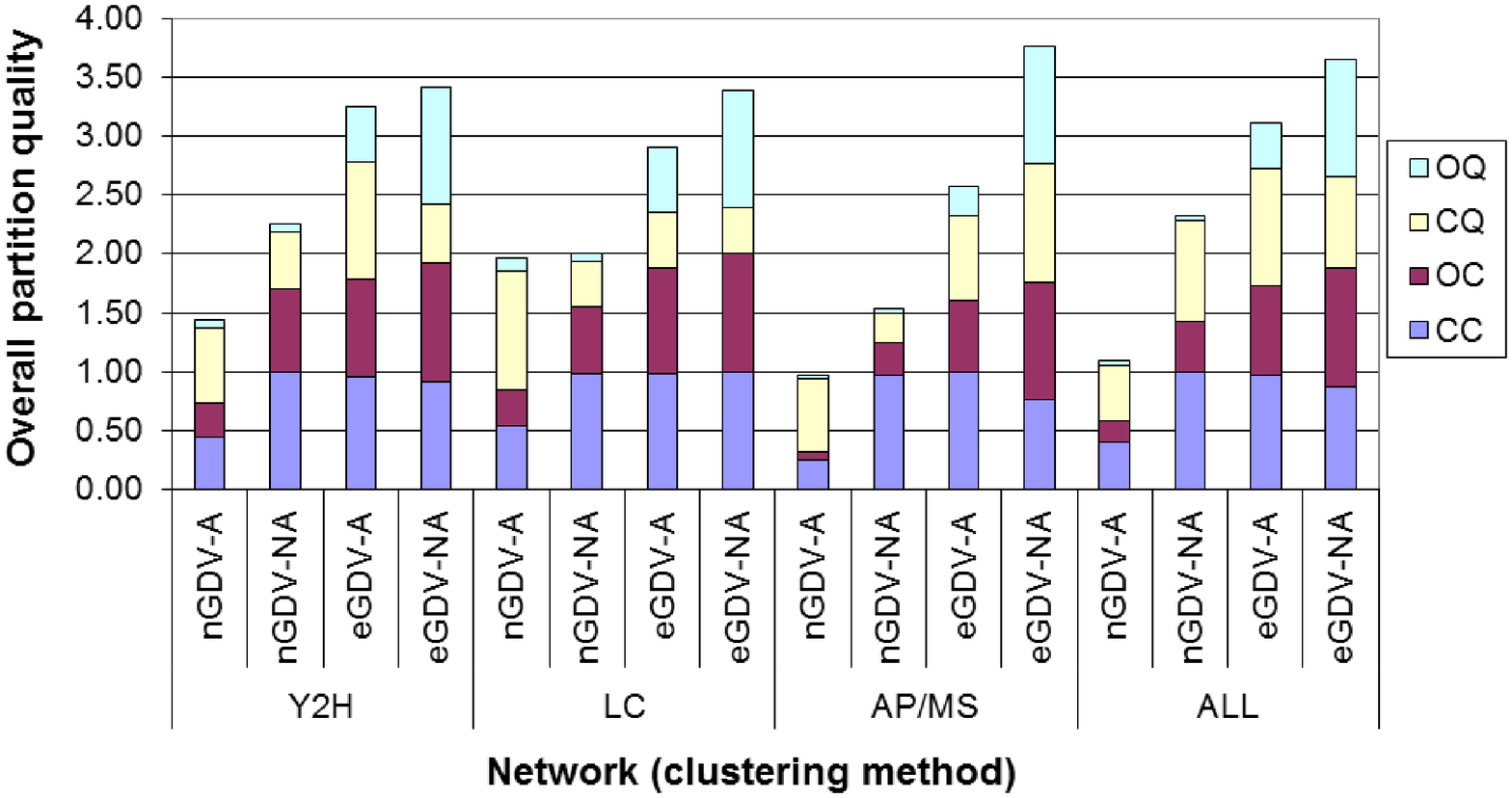}}
\caption{Comparison of our node and edge clustering
  methods on the four yeast PPI networks (Y2H, LC, AP/MS, and
  ALL). nGDV-A denotes the node clustering method when
  node-GDV-similarity is used as the distance metric for hierarchical
  clustering of adjacent nodes only and the partition with the maximum
  partition density is selected.  nGDV-NA denotes the node clustering
  method when node-GDV-similarity is used as the distance metric for
  hierarchical clustering of both adjacent and non-adjacent nodes and
  the partition with the best overall partition quality is selected.
  eGDV-A denotes the edge clustering method when edge-GDV-similarity
  is used as the distance metric for hierarchical clustering of
  adjacent edges only and the partition with the maximum partition
  density is selected. eGDV-NA denotes the edge clustering method when
  edge-GDV-similarity is used as the distance metric for hierarchical
  clustering of both adjacent and non-adjacent edges and the partition
  with the best overall partition quality is selected.  The clustering
  methods are compared with respect to the following measures: cluster
  coverage (CC), overlap coverage (OC), cluster quality (CQ), and
  overlap quality (OQ). The overall partition quality score ($y$-axis)
  is the sum of these four measures after each is normalized to [0,1],
  such that the maximum achievable score is four. We compare our
  approach when using edge-GDV-similarity as the distance metric for
  edge clustering against our approach when using node-GDV-similarity
  as the distance metric for node clustering since we want to answer
  if and how much we gain by clustering of edges compared to
  clustering of nodes. And to answer this, one should use conceptually
  similar edge and node clustering methods, such as these. The figure
  shows that in each network: 1) we gain by clustering both adjacent
  and non-adjacent nodes compared to clustering only adjacent nodes;
  2) we gain further by clustering adjacent edges instead of
  clustering nodes; and 3) we gain the most by clustering both
  adjacent and non-adjacent edges. }  \label{fig:nGDV_eGDV}
\end{center} 
\end{figure*}

We gain by using edge-GDV-similarity for clustering: eGDV-A-D
outperforms all node clustering approaches on all networks. (This
includes node clustering by using node-GDV-similarity, as shown in
Fig. \ref{fig:nGDV_eGDV}). Also, it outperforms edge-SN on Y2H and AP/MS.
Although edge-SN is slightly better than and comparable to eGDV-A-D on
LC and ALL networks, respectively, eGDV-NA-D outperforms edge-SN on
these two networks, as well as on AP/MS.  Hence, we gain further by
clustering non-adjacent edges in addition to adjacent ones. The only
exception is Y2H, for which edge-SN is slightly better than eGDV-NA-D.
However, as already noted, eGDV-A-D outperforms edge-SN on Y2H
network.  Hence, we are always superior, with either eGDV-A-D or
eGDV-NA-D or both eGDV-A-D and eGDV-NA-D.  With eGDV-NA-B, we further
demonstrate our superiority over all other methods on all networks.

\end{document}